# An optimal dispatch scheme for DSO and prosumers by implementing three-phase distribution locational marginal prices

Jiaqi Chen [1], Ye Guo [1], Wenchuan Wu [1,2*]

[1] Tsinghua-Berkeley Shenzhen Institute, Tsinghua University, Shenzhen, People's Republic of China
[2] State Key Laboratory of Power Systems, Department of Electrical Engineering, Tsinghua University, Beijing, People's Republic of China
* wuwench@tsinghua.edu.cn

**Abstract:** Since distribution system operator (DSO) cannot directly control prosumers with controllable resources, this paper proposes an optimal dispatch method of using three-phase distribution locational marginal prices (DLMPs) as effective economic signals to incentivize prosumers' behaviors. In the proposed three-phase DLMP model, DLMPs for both active power and reactive power are calculated. To alleviate the imbalance, congestions and voltage violations in active distribution networks (ADNs), the DSO and prosumers should be coordinated. We develop such a coordinated control scheme for the DSO and prosumers, in which the DSO generates and broadcasts three-phase DLMPs as price signals to induce prosumers' behaviors. We prove that given the DLMPs for active power and reactive power as settlement prices, the optimal dispatch of the ADN will also maximize the surplus of prosumers. Therefore, the power output of rational prosumers will match the optimal dispatch, resulting in better operational conditions of ADNs. Then the three-phase imbalance, congestions and voltage violations will be well reduced. Numerical tests demonstrate the effectiveness of the proposed approach.

## 1. Introduction

### 1.1. Background

With the increasing amount of distributed generations (DGs), power end-users are transforming from traditional passive consumers to prosumers that also actively provide energy services to the grid [1]. The operation of active distribution networks (ADNs) has encountered significant challenges because of the large penetration of DGs and flexible loads (FLs) in prosumers. These active resources may make the state of the distribution network more volatile, and may even lead to power overflows on distribution lines or transformer windings and voltage over-limits. Moreover, ADNs are typically three-phase imbalanced systems. If its degree of three-phase imbalance is too high, the AND may suffer from additional power loss, congestions on distribution lines and transformers and voltage deviations of neutral points [2]. To handle these operational issues adequately, the distribution system operator (DSO) should coordinate with proactive prosumers in its territory to make use of their controllable resources. To better integrate the DGs and FLs, a new energy dispatch framework for active distribution network should be established, with a proper market design that can maximize the social welfare and ensure the legal interests of market participants [3].

### 1.2. Literature review

There are many direct control methods such as shedding loads directly by system operators to address the operational issues [4]-[5], but these methods neglect the impact from proactive customers, which is inadequate for the operation of ADNs. The distribution locational marginal price (DLMP) is developed in papers [6]-[8], reflecting the marginal cost of supplying the next incremental loads at different locations.

DLMP can be used as economic signals to incentivize prosumers to optimally adjust their power consumptions and generations according to its physical significance. Papers [9]-[11] use DLMP in day-ahead markets to alleviate congestions in future distribution networks with high penetrations of EVs. To lift congestions with demand response, a distribution congestion price based on DLMP is proposed in [12]. In [13], DLMP is calculated for both active power and reactive power to motivate distributed energy resources to contribute to congestion management and voltage support, based on a mixed-integer second order-cone programming (SOCP) model for DSO optimization considering network losses. The above applications of DLMP are proven to be effective in congestion and voltage management. However, they are still based on single-phase models without considering the imbalance characteristic of ADNs.

To take the imbalance characteristic in ADNs into account, a three-phase DLMP based on the AC optimal power flow (ACOPF) model is presented in [14]. Therein, three-phase DLMP is modeled as the marginal cost to serve an incremental unit of demand at a specific phase at a certain node, and is calculated using the Lagrangian multipliers in the three-phase ACOPF model. The optimal Lagrangian multipliers can accurately reflect the sensitivity of the optimal cost with respect to changes in bounds of the constraints on demands at the global optimal solution. However, the ACOPF problem is nonconvex and it is hard to solve for the global optimal solution. To transform the ACOPF problem into a convex problem, SOCP relaxation is presented in [15]-[18] and is used to make the single-phase ACOPF problem convex to calculate DLMP [13]. But SOCP relaxation can only be applied in single-phase model [19]. So the three-phase DLMP cannot be achieved by solving three-phase ACOPF model using SOCP relaxation.

To solve this problem, paper [20] applies a sparse moment relaxation approach to solve the three-phase ACOPF problem and to calculate the three-phase DLMP. But the optimal solution to the moment relaxation problem may not be feasible to the original ACOPF especially for low order moment relaxations. In addition, this method may suffer from poor computation efficiencies so may not satisfy real time DLMP calculation in large scale ADNs. Therefore, it is necessary to find a method to calculate three-phase DLMP efficiently and accurately.

*1.3. Contributions*

In order to reduce three-phase imbalance, line congestions and nodal voltage violations in ADNs, a market based optimal dispatch method by implementing three-phase DLMP is proposed in this paper. Comparing to existing literature, the optimal dispatch model of ADNs adopted in this paper has the following characteristics: (i) A new constraint on the level of three-phase imbalance has been added, which is measured by the NEMA (National Electric Manufacturers Associations of the USA) Std; (ii) To calculate three-phase DLMPs efficiently, we employ a linearized power flow model method to make the three-phase ACOPF model convex; (iii) DLMPs for both active and reactive powers have been derived. By solving such an optimal dispatch problem, the DSO obtains the optimal quantities and prices.

However, the DSO cannot really mandate the power output of DGs owned by prosumers. Instead, the DSO broadcasts three-phase prices to prosumers and expects them to response accordingly. Thus the crux of our problem is whether prosumers will response "reasonably" or not. This paper partly answers this question by proving that if the DSO broadcasts three-phase DLMPs to prosumers, then the prosumers' optimal responses that maximize their surpluses are generating/consuming as per the optimal quantities solved from the optimal dispatch problem. Therefore, if all prosumers are rational and follow the optimal dispatch results, the security and efficiency of ADNs will be guaranteed.

The remainder of this paper is organized as follows. Section 2 describes the optimization models for DSO and prosumers, and determines the three-phase DLMP for active power and reactive power. Section 3 introduces the method that using DLMP to alleviate three-phase imbalance, congestions and voltage violations in ADNs, and gives the proof process. Section 4 uses numerical simulations to analyses the effectiveness of the proposed method, and conclusions are drawn in Section 5.

## 2. Problem modelling

*2.1. OPF model of DSO with imbalance constraint*

In this paper, the degree of three-phase imbalance of a distribution network is measured by the following index $\delta_i$ according to the standard of National Electric Manufacturers Associations [21].

$$\delta_i = \frac{\max_{\varphi=a,b,c} |V_i^\varphi - V_i^{av}|}{V_i^{av}}, \quad V_i^{av} = \frac{\sum_{\varphi=a,b,c} V_i^\varphi}{3}, \quad (1)$$

where parameters $V_i^\varphi, V_i^a, V_i^b, V_i^c$ are voltage magnitudes at node $i$ at phase $\varphi, a, b, c$, parameter $V_i^{av}$ is the average of three-phase voltage magnitudes at node $i$.

Accordingly, we model the limit on the degree of three-phase imbalances as operational constraints, together with limits on branch power flows and nodal voltage magnitudes. Moreover, we adopt the linearized distribution power flow model [22] to make the OPF problem convex. The resulting optimal dispatch model is as follows:

$$(P1) \max_{\{p_d, p_g, q_g, p_b, q_b, u, \delta_i\}} U(p_d) - C(p_g), \quad (2)$$

$$s.t. \; \boldsymbol{p_b} = \boldsymbol{K_1} \cdot \left[ (\boldsymbol{p_g} - \boldsymbol{p_d}) + \boldsymbol{K_2} \cdot (\boldsymbol{q_g} - \boldsymbol{q_d}) + \boldsymbol{K_3} \right], \quad (3)$$

$$\boldsymbol{q_b} = \boldsymbol{K_4} \cdot \left[ (\boldsymbol{q_g} - \boldsymbol{q_d}) + \boldsymbol{K_5} \cdot (\boldsymbol{p_g} - \boldsymbol{p_d}) + \boldsymbol{K_6} \right], \quad (4)$$

$$\boldsymbol{u} = \boldsymbol{K_7} \cdot \boldsymbol{p_b} + \boldsymbol{K_8} \cdot \boldsymbol{q_b} + \boldsymbol{K_9}, \quad (5)$$

$$u_{root}^\varphi = u_{ref}^\varphi, \quad (6)$$

$$p_{root}^\varphi = \sum_{l \in \mathcal{L}_{root}} p_{b,l}^\varphi, \quad (7)$$

$$q_{root}^\varphi = \sum_{l \in \mathcal{L}_{root}} q_{b,l}^\varphi, \quad (8)$$

$$\underline{\boldsymbol{p_b}} \leq \boldsymbol{p_b} \leq \overline{\boldsymbol{p_b}}, \quad (9)$$

$$\underline{\boldsymbol{u}} \leq \boldsymbol{u} \leq \overline{\boldsymbol{u}}, \quad (10)$$

$$\underline{\boldsymbol{p_g}} \leq \boldsymbol{p_g} \leq \overline{\boldsymbol{p_g}}, \quad (11)$$

$$\underline{\boldsymbol{p_d}} \leq \boldsymbol{p_d} \leq \overline{\boldsymbol{p_d}}, \quad (12)$$

$$\underline{\boldsymbol{q_g}} \leq \boldsymbol{q_g} \leq \overline{\boldsymbol{q_g}}, \quad (13)$$

$$\delta_i \leq \overline{\delta}, \forall i \in \mathcal{N}, \quad (14)$$

where the objective function (2) is to maximize the social welfare as the difference between the total utility of prosumers $U(\boldsymbol{p_d})$ and the total cost of DGs and electricity purchased from the main grid $C(\boldsymbol{p_g})$. Vectors $\boldsymbol{p_d}, \boldsymbol{p_g}$ are the active power demand and generators' active power output at all nodes, respectively. In this model, we assume the root node is the power supply point (PSP), which can be regarded as a generator with infinite capacity to balance the demand and supply.

Constraints (3)-(5) represent branch power flows as functions of power injections, which are derived from linearized power flow equations, where vectors $\boldsymbol{p_b}, \boldsymbol{q_b}$ are three-phase branch active and reactive power flows, respectively, and vectors $\boldsymbol{q_d}, \boldsymbol{q_g}$ are three-phase reactive power demand and generators' reactive power output, vector $\boldsymbol{u}$ is the square of three-phase voltage magnitude, coefficient matrices $\boldsymbol{K_1}$ to $\boldsymbol{K_9}$ are constant. More details about the three-phase linear power flow model can be found in paper [22]. We assume that the root node is the slack bus, with the voltage constraint (6), where $u_{root}^\varphi$ is squared voltage magnitude at root node at phase $\varphi$, $u_{ref}^\varphi$ is the square of reference voltage magnitude. Constraints (7)-(8) are power balance constraints, representing the active and reactive power at root node should be equal to the sum of active and reactive power of the branches that connected to the root node, respectively, where $\mathcal{L}_{root}$ is the set of branches that connected to the root node. Constraints (9)-(13) represent, respectively, lower and upper limits of branch active power $\underline{p}_b, \overline{p}_b$, voltage magnitude $\underline{u}, \overline{u}$, generators' active power output $\underline{p}_g, \overline{p}_g$, active power demand $\underline{p}_d, \overline{p}_d$ and generators' reactive power output $\underline{q}_g, \overline{q}_g$. Because prosumers always make response in active power, in this model, we assume reactive power demand $q_d$ is constant. We also incorporate the limit on three-phase imbalance (14) into the optimal dispatch model. Constraint (14) represents the imbalance index $\delta_i$ should not exceed its limit $\overline{\delta}$, and $\mathcal{N}$ is the set of all nodes. To maintain the convexity of the problem, we approximate (14) using (15), where parameter $u_i^\varphi$ is squared voltage magnitude at node $i$ at phase $\varphi$. The derivation of the approximate process is introduced in Appendix.

$$(1-\overline{\delta})^2 \sum u_i^\varphi \leq 3u_i^\varphi \leq (\overline{\delta}+1)^2 \sum u_i^\varphi, \varphi = a,b,c. \quad (15)$$

Here we assume that, the price to buy electricity from the main grid is the locational marginal price (LMP) $\pi_{LMP}$ at the root node, which is constant, so the corresponding cost function $C(p_{g,root}^\varphi)$ for root node at phase $\varphi$ can be expressed as (16),

$$C(p_{g,root}^\varphi) = \pi_{LMP} \cdot p_{g,root}^\varphi, \quad (16)$$

where $p_{g,root}^\varphi$ is the active power of root node injection at phase $\varphi$.

For each prosumer and generator $i$ in phase $\varphi$, its utility function $U_i^\varphi(p_{d,i}^\varphi)$ and cost function $C_i^\varphi(p_{g,i}^\varphi)$ are quadratic [23]-[25], which are written as (17)-(18),

$$U_i^\varphi(p_{d,i}^\varphi) = c_{1,i}^\varphi (p_{d,i}^\varphi)^2 + c_{2,i}^\varphi p_{d,i}^\varphi + c_{3,i}^\varphi, \quad (17)$$

$$C_i^\varphi(p_{g,i}^\varphi) = c_{4,i}^\varphi (p_{g,i}^\varphi)^2 + c_{5,i}^\varphi p_{g,i}^\varphi + c_{6,i}^\varphi, \quad (18)$$

where parameters $c_{1,i}^\varphi$ to $c_{6,i}^\varphi$ are coefficients of the corresponding functions.

In this way, the optimal dispatch problem of DSO is a convex QP problem with linear constraints, which can be summarize into the following compact form.

$$(P2) \max_{\{p_d, p_g, q_g\}} U(p_d) - C(p_g), \quad (19)$$

$$s.t. \quad M \cdot x + m \leq 0, (\mu_M), \quad (20)$$

$$N \cdot x + n = 0, (\lambda_N), \quad (21)$$

$$\underline{x} \leq x \leq \overline{x}, (\mu_x^-, \mu_x^+), \quad (22)$$

where

$$x = \begin{bmatrix} p_d & p_g & q_g \end{bmatrix}^T, M = \begin{bmatrix} M_{pd} & -M_{pd} & M_{qg} \end{bmatrix},$$
$$N = \begin{bmatrix} N_{pd} & -N_{pd} & N_{qg} \end{bmatrix}, \mu_x^- = \begin{bmatrix} \mu_{pd}^- & \mu_{pg}^- & \mu_{qg}^- \end{bmatrix}^T, \quad (23)$$
$$\mu_x^+ = \begin{bmatrix} \mu_{pd}^+ & \mu_{pg}^+ & \mu_{qg}^+ \end{bmatrix}^T.$$

Inequality constraint (20) is derived from constraints (3)-(5), (9)-(10) and (15), equality constraint (21) is derived from constraints (3)-(8), inequality constraint (22) represents constraints (11)-(13). The vector $x$ consists of three-phase power demand $p_d$, real power generation $p_g$, and reactive power generation $q_g$, with the lower and upper limit $\underline{x}, \overline{x}$. Coefficient matrixes $M, N$ include submatrices $M_{pd}, -M_{pd}, M_{qg}$ and $N_{pd}, -N_{pd}, N_{qg}$ corresponding to parameters $p_g, p_g, q_g$, respectively. Parameter vectors $m, n$ are constant. Vectors $\mu_M, \lambda_N, \mu_x^-, \mu_x^+$ are dual variables of corresponding constraints. The superscript $T$ denotes transpose.

### 2.2. Derivation of Three-phase DLMP

In ADNs, real and reactive powers are closely correlated, and it is necessary to consider the impact of reactive power in the dispatch problem. Therefore, we derive the three-phase DLMP for reactive power, referred to as Q-DLMP, which represents the incremental cost to supply an extra unit of reactive power demand at a specific phase at a certain bus. Since the optimal dispatch model of DSO is convex with zero duality gap, we can construct the three-phase DLMP via Lagrangian multipliers of corresponding constraints. The Lagrangian of the optimal dispatch problem P2 is expressed as follows,

$$\begin{aligned} L(x, \mu_M, \lambda_N, \mu_x^-, \mu_x^+) &= -U(p_d) + C(p_g) \\ &+ \mu_M^T \cdot (M_{pd} \cdot p_d - M_{pd} \cdot p_g + M_{qg} \cdot q_g + m) \\ &+ \lambda_N^T \cdot (N_{pd} \cdot p_d - N_{pd} \cdot p_g + N_{qg} \cdot q_g + n) \\ &+ \mu_x^{+T} \cdot (x - \overline{x}) - \mu_x^{-T} \cdot (x - \underline{x}) \end{aligned} \quad (24)$$

Assume the considered OPF problem has an optimal solution $(x^*)$, and $(\mu_M^*, \lambda_N^*, \mu_x^{+*}, \mu_x^{-*})$ are the corresponding Lagrangian multipliers, the three-phase active power DLMP $\pi_{DLMP}^p$, referred to as P-DLMP, can be derived respect to the inelastic active demand when the problem reaches its optimum. The corresponding Lagrangian with inelastic active demand $p_d$ is as follows,

$$\begin{aligned} L(x^*, \mu_M^*, \lambda_N^*, \mu_x^{+*}, \mu_x^{-*}) &= -U(p_d^*) + C(p_g^*) \\ &+ (\mu_M^*)^T \cdot [M_{pd} \cdot (p_d^* + p_d) - M_{pd} \cdot p_g^* + M_{qg} \cdot q_g^* + m] \\ &+ (\lambda_N^*)^T \cdot [N_{pd} \cdot (p_d^* + p_d) - N_{pd} \cdot p_g^* + N_{qg} \cdot q_g^* + n] \\ &+ (\mu_x^{+*})^T \cdot (x^* - \overline{x}) - (\mu_x^{-*})^T \cdot (x^* - \underline{x}) \end{aligned} .(25)$$

So the P-DLMP can be derived as follows,

$$\pi_{DLMP}^p = \left.\frac{\partial(-f)}{\partial p_d}\right|_{x^*} = \left.\frac{\partial L}{\partial p_d}\right|_{x^*, \mu_M^*, \lambda_N^*, \mu_x^{+*}, \mu_x^{-*}}, \quad (26)$$
$$= M_{pd}^T \cdot \mu_M^* + N_{pd}^T \cdot \lambda_N^*$$

where $f$ represents the objective function (19) of P2.

For reactive power, the change of demand $q_d$ can be regarded as the opposite change of generators' output $q_g$, so the three-phase Q-DLMP $\pi_{DLMP}^q$ can be derived in a similar way as follows,

$$\pi_{DLMP}^{q} = \frac{\partial(-f)}{\partial \boldsymbol{q}_d}\bigg|_{x^*} = \frac{\partial L}{\partial \boldsymbol{q}_d}\bigg|_{x^*,\mu_M^*,\lambda_N^*,\mu_x^{+*},\mu_x^{-*}}. \quad (27)$$
$$= -\boldsymbol{M}_{qg}^T \cdot \boldsymbol{\mu}_M^* - \boldsymbol{N}_{qg}^T \cdot \boldsymbol{\lambda}_N^*$$

In the optimal dispatch model for DSO, we consider the power loss, line flow constraints, voltage magnitude constraints and three-phase voltage imbalance constrains, so the value of proposed three-phase P-DLMP and Q-DLMP also consist of components reflecting the power loss, congestions, as well as voltage and imbalance level limits. In what follows, we demonstrate that taking the proposed three-phase DLMP as price signals, the branch power overflows, voltage violations and three-phase imbalance will be alleviated.

### 2.3. Response of prosumers

In this paper, we assume that prosumers are price takers, and the active power demand and DGs' active and reactive output of prosumers are elastic with respect to prices. Prosumers are assumed to be economically rational and they will make response to the price signal to maximize their own surplus. The optimization model for prosumer $i$ at phase $\varphi$ is as follows.

$$(\text{P3}) \max_{\{p_{d,i}^{\varphi}, p_{g,i}^{\varphi}, q_{g,i}^{\varphi}\}} U_i^{\varphi}(p_{d,i}^{\varphi}) + \pi_i^{p,\varphi} \cdot p_{g,i}^{\varphi} + \pi_i^{q,\varphi} \cdot q_{g,i}^{\varphi} \\ - C_i^{\varphi}(p_{g,i}^{\varphi}) - \pi_i^{p,\varphi} \cdot p_{d,i}^{\varphi} \quad , \quad (28)$$

$$s.t. \quad \underline{x}_i^{\varphi} \leq x_i^{\varphi} \leq \overline{x}_i^{\varphi}, (\mu_{x,i}^{\varphi-}, \mu_{x,i}^{\varphi+}) \quad , \quad (29)$$

where

$$x_i^{\varphi} = \begin{bmatrix} p_{d,i}^{\varphi} & p_{g,i}^{\varphi} & q_{g,i}^{\varphi} \end{bmatrix}^T, \mu_{x,i}^{\varphi-} = \begin{bmatrix} \mu_{pd,i}^{\varphi-} & \mu_{pg,i}^{\varphi-} & \mu_{qg,i}^{\varphi-} \end{bmatrix}^T, \\ \mu_{x,i}^{\varphi+} = \begin{bmatrix} \mu_{pd,i}^{\varphi+} & \mu_{pg,i}^{\varphi+} & \mu_{qg,i}^{\varphi+} \end{bmatrix}^T. \quad (30)$$

The objective of prosumers is to maximize their individual surplus, which is expressed as the demand utility $U_i^{\varphi}(p_{d,i}^{\varphi})$ plus the benefit to sell the DGs' output $(\pi_i^{p,\varphi} \cdot p_{g,i}^{\varphi} + \pi_i^{q,\varphi} \cdot q_{g,i}^{\varphi})$ minus DGs' cost $C_i^{\varphi}(p_{g,i}^{\varphi})$ and the payment to buy electricity from main grid $\pi_i^{p,\varphi} \cdot p_{d,i}^{\varphi}$. Parameters $\pi_i^{p,\varphi}, \pi_i^{q,\varphi}$ are prices for active and reactive power that prosumer $i$ at phase $\varphi$ is charged, parameter $q_{g,i}^{\varphi}$ is generators' reactive output at node $i$ at phase $\varphi$, parameter $x_i^{\varphi}$ represents the variables, include active power demand $p_{d,i}^{\varphi}$, generators' active power output $p_{g,i}^{\varphi}$ and generators' reactive power output $q_{g,i}^{\varphi}$. Prosumers only knows the charging price and their own information, so the constraints are the lower and upper bounds $\underline{x}_i^{\varphi}, \overline{x}_i^{\varphi}$ of the demand and generators' output (29), without any network constraints and operation constraints. Parameters $\mu_{x,i}^{\varphi-}, \mu_{x,i}^{\varphi+}$ are dual variables of corresponding constraints.

## 3. Three-phase DLMP based optimal dispatch method

In a real electricity market, DSO cannot really control prosumers and their DGs directly, so releasing price signals and anticipating prosumers to react accordingly is the best possible way to manage the system. In the current electricity market, different prosumers are charged a given flat tariff which is often set as the LMP at the root node. Charging different prosumers the same price is not fair, and the given flat tariff will not reflect the difference of three phases, power loss and congestions in ADNs, or control the operation of the system, so we use the proposed three-phase P-DLMP and Q-DLMP to guide the prosumers' behavior relating to active and reactive power, respectively, and then the branch power overflows, voltage violations and three-phase imbalance will be alleviated.

We have shown in the following theorem and corollary that taking the proposed three-phase DLMP as price signals, the active demand power as well as the generators' active and reactive output power will obey the dispatch of DSO.

**Theorem 1**: Assume that the settlement prices $\pi_i^{p,\varphi}, \pi_i^{q,\varphi}$ are determined by (26) and (27) as P-DLMP and Q-DLMP, then the optimal power consumption $p_{d,i}^{\varphi}*$ and the output of DGs $p_{g,i}^{\varphi}*, q_{g,i}^{\varphi}*$ of an arbitrary prosumer $i$ in the optimal dispatch model P2 will also be optimal for the surplus maximization model P3 for the same prosumer.

*Proof:*

The KKT conditions of the DSO's optimal dispatch problem P2 are

$$\begin{aligned} &-U'(\boldsymbol{p}_d) + \boldsymbol{M}_{pd}^T \cdot \boldsymbol{\mu}_M + \boldsymbol{N}_{pd}^T \cdot \boldsymbol{\lambda}_N + \boldsymbol{\mu}_{pd}^+ - \boldsymbol{\mu}_{pd}^- = 0, \\ &C'(\boldsymbol{p}_g) - \boldsymbol{M}_{pd}^T \cdot \boldsymbol{\mu}_M - \boldsymbol{N}_{pd}^T \cdot \boldsymbol{\lambda}_N + \boldsymbol{\mu}_{pg}^+ - \boldsymbol{\mu}_{pg}^- = 0, \\ &\boldsymbol{M}_{qg}^T \cdot \boldsymbol{\mu}_M + \boldsymbol{N}_{qg}^T \cdot \boldsymbol{\lambda}_N + \boldsymbol{\mu}_{qg}^+ - \boldsymbol{\mu}_{qg}^- = 0, \\ &\boldsymbol{N} \cdot \boldsymbol{x} + \boldsymbol{n} = 0, \\ &\boldsymbol{\mu}_M \cdot (-\boldsymbol{M} \cdot \boldsymbol{x} - \boldsymbol{m}) = 0, \\ &\boldsymbol{\mu}_x^- \cdot (\boldsymbol{x} - \underline{\boldsymbol{x}}) = 0, \\ &\boldsymbol{\mu}_x^+ \cdot (-\boldsymbol{x} + \overline{\boldsymbol{x}}) = 0, \\ &\boldsymbol{\mu}_M \geq 0, \ -\boldsymbol{M} \cdot \boldsymbol{x} - \boldsymbol{m} \geq 0, \\ &\boldsymbol{\mu}_x^- \geq 0, \ \boldsymbol{x} - \underline{\boldsymbol{x}} \geq 0, \\ &\boldsymbol{\mu}_x^+ \geq 0, \ -\boldsymbol{x} + \overline{\boldsymbol{x}} \geq 0, \end{aligned} \quad (31)$$

where parameter $U'(\boldsymbol{p}_d)$ is the derivative of utility function respect to prosumers' active demand $\boldsymbol{p}_d$, parameter $C'(\boldsymbol{p}_g)$ is the derivative of cost function respect to generators' active output $\boldsymbol{p}_g$.

The KKT conditions of the prosumers' problem P3 are

$$\begin{aligned} &-U_i^{\varphi'}(p_{d,i}^{\varphi}) + \pi_i^{p,\varphi} + \mu_{pd,i}^{\varphi+} - \mu_{pd,i}^{\varphi-} = 0, \\ &C_i^{\varphi'}(p_{g,i}^{\varphi}) - \pi_i^{p,\varphi} + \mu_{pg,i}^{\varphi+} - \mu_{pg,i}^{\varphi-} = 0, \\ &-\pi_i^{q,\varphi} + \mu_{qg,i}^{\varphi+} - \mu_{qg,i}^{\varphi-} = 0, \\ &\mu_{x,i}^{\varphi-} \cdot (x_i^{\varphi} - \underline{x}_i^{\varphi}) = 0, \\ &\mu_{x,i}^{\varphi+} \cdot (-x_i^{\varphi} + \overline{x}_i^{\varphi}) = 0, \\ &\mu_{x,i}^{\varphi-} \geq 0, \ x_i^{\varphi} - \underline{x}_i^{\varphi} \geq 0, \\ &\mu_{x,i}^{\varphi+} \geq 0, \ -x_i^{\varphi} + \overline{x}_i^{\varphi} \geq 0, \end{aligned} \quad (32)$$

$$(33)$$

where parameters $U_i^{\varphi'}(p_{d,i}^{\varphi}), C_i^{\varphi'}(p_{g,i}^{\varphi})$ are the derivative of corresponding functions for prosumer $i$ at phase $\varphi$.

In the proposed method, the prices $\pi_i^{p,\varphi}, \pi_i^{q,\varphi}$ sent to prosumers are P-DLMP $\pi_{DLMP}^{p,\varphi}$ and Q-DLMP $\pi_{DLMP}^{q,\varphi}$, of which the values are denoted in (26)-(27), assuming that $\left(p_d*, p_g*, q_g*, \mu_M*, \lambda_N*, \mu_x^+*, \mu_x^-*\right)$ is a solution to the KKT conditions of the DSO's problem P2, so equations (32) will be transformed into

$$-U_i^{\varphi}{}'(p_{d,i}^{\varphi}) + \left(\boldsymbol{M}_{pd}^T \cdot \boldsymbol{\mu}_M* + \boldsymbol{N}_{pd}^T \cdot \boldsymbol{\lambda}_N*\right)_i^{\varphi} + \mu_{pd,i}^{\varphi+} - \mu_{pd,i}^{\varphi-} = 0,$$

$$C_i^{\varphi}{}'(p_{g,i}^{\varphi}) - \left(\boldsymbol{M}_{pd}^T \cdot \boldsymbol{\mu}_M* + \boldsymbol{N}_{pd}^T \cdot \boldsymbol{\lambda}_N*\right)_i^{\varphi} + \mu_{pg,i}^{\varphi+} - \mu_{pg,i}^{\varphi-} = 0, \quad (34)$$

$$\left(\boldsymbol{M}_{qg}^T \cdot \boldsymbol{\mu}_M* + \boldsymbol{N}_{qg}^T \cdot \boldsymbol{\lambda}_N*\right)_i^{\varphi} + \mu_{qg,i}^{\varphi+} - \mu_{qg,i}^{\varphi-} = 0,$$

where $(\cdot)_i^{\varphi}$ represents the element at node $i$ at phase $\varphi$.

By comparing the KKT conditions, it is evident that $\left(p_{d,i}^{\varphi}*, p_{g,i}^{\varphi}*, q_{g,i}^{\varphi}*, \mu_{x,i}^{\varphi+}*, \mu_{x,i}^{\varphi-}*\right)$, which is the solution to the KKT conditions of the DSO's problem P2 at node $i$ of phase $\varphi$, satisfies the KKT conditions (33) and (34) of the prosumers' problem P3. This means $\left(p_{d,i}^{\varphi}*, p_{g,i}^{\varphi}*, q_{g,i}^{\varphi}*\right)$, which is the optimal solution to DSO problem P2 at node $i$ at phase $\varphi$, is also an optimal solution to the prosumer's problem P3. □

**Corollary 1:** Under the same assumption as in **Theorem 1** and when Q-DLMP is not equal to zero, then the optimal solution to the prosumer in P3 is unique and it is equal to that solved from the optimal dispatch problem P2.

*Proof:*

The objective function (28) of P3 have quadratic functions respect to active power $p_{d,i}^{\varphi}, p_{g,i}^{\varphi}$ with positive definite Hessian matrix, and the constraints (29) are affine functions, so the prosumers' optimization problem are strictly convex QP problem respect to active power. So the active power optimal solution is unique. Because the prosumers' objective function (28) has an affine form respect to reactive power, the reactive power optimal solution will reach the upper bound of reactive power limit if price $\pi_i^{q,\varphi}$ is positive, and will reach the lower bound if price $\pi_i^{q,\varphi}$ is negative. When we use P-DLMP and Q-DLMP as price signals and Q-DLMP is not equal to zero, the solution to prosumers' problem is unique. According to **Theorem 1** that the optimal solution to DSO problem P2 at node $i$ at phase $\varphi$ is also a solution to the prosumer's problem P3, and because of the uniqueness of the optimal solution to the prosumers' problem P3, any optimal solution to the prosumers' problem must also be the optimal solution to the DSO problem at node $i$ at phase $\varphi$. Based on above conclusions, the DSO problem P2 and the prosumers' problem P3 are equivalent. □

Consequently, prosumers will make demand response to DLMP to maximize their surplus, at the same time social welfare will be maximized. In addition, in practice, DGs' reactive outputs will help to reduce active power loss in the system, because the prosumers' DGs' output range is small compared with whole reactive power demands, the reactive outputs of DGs will usually reach their bounds. In this case, the corresponding dual variables $\mu_{qg,i}^{\varphi+}$ or $\mu_{qg,i}^{\varphi-}$ are nonzero. According to equation (35), the calculated Q-DLMPs will usually not equal to zero, so the proposed method can be applied in most cases. And if the Q-DLMPs are equal to zero, the dispatch of prosumers will remain unchanged.

$$\left(\boldsymbol{M}_{qg}^T \cdot \boldsymbol{\mu}_M* + \boldsymbol{N}_{qg}^T \cdot \boldsymbol{\lambda}_N*\right)_i^{\varphi} + \mu_{qg,i}^{\varphi+} - \mu_{qg,i}^{\varphi-} = 0 \quad . \quad (35)$$

In particular, this proved equivalence is not a trivial extension of the results of the LMP in wholesale market in that: (i) the prosumer's optimal behavior under nonzero Q-DLMP in our proposed method is unique; (ii) both reactive power and voltage constraints are taken into account in our method, but the wholesale market and LMP only focus on the active power.

In the proposed scheme, DSO solves the optimal dispatch problem P2 at the start of each price interval, then calculates the three-phase DLMPs and broadcasts the price. Prosumers will solve their own optimizations P3, and under given assumptions, they will voluntarily follow the instructions of the DSO. The framework of the proposed method is illustrated in Fig.1.

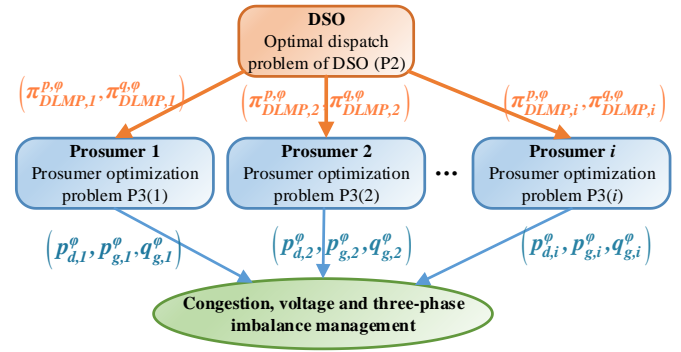

**Fig. 1** *Illustrate of framework of proposed optimal dispatch method*

Because the two optimal problems P2, P3 are equivalent, no iteration is needed in the process. Note that in the optimal dispatch problem P1, voltages and branch power flows are in acceptable ranges and the degree of three-phase imbalance is less than the limit $\bar{\delta}$. Therefore, if prosumers follow their dispatch instructions, the three-phase imbalance, branch power flow and voltage will be controlled within the expect limitations.

In practice, forecast error and uncertainty of prosumers may cause the branch power overflows, voltage violations and three-phase imbalance although we use the proposed method. But still, compared with current electricity market mechanism that sends prosumers the flat tariff signals, which is often set as the LMP at the root node, the proposed method will control imbalance, congestions and voltage better. The more accurate of the forecast and the more stable of prosumers, the more significant the three-phase DLMP will manage the operation of the system. We have also demonstrated this in our simulations in the next section.

## 4. Numerical tests

### 4.1. Simulation setup

We tested the effectiveness of three-phase DLMP in alleviating branch power overflows, voltage violations and three-phase imbalance on a modified three-phase IEEE 33-bus distribution system shown as Fig.2. More information about branch parameters and load profiles is available online [26]. Different prosumers at different phases and nodes had different utility functions, which made the system imbalanced. Node 1 was the PSP, which was regarded as a conventional generator with infinite capacity. Six different DGs (DG1, DG2, DG3, DG4, DG5, DG6) were connected to Bus 3, 6, 12, 18, 22, and 33 separately, and the active output limit of DGs was [0, 0.1] MW, the LMP at PSP was $30/MWh. The coefficients of DGs' cost functions were listed in Table 1. The output of DGs was assumed phase independent.

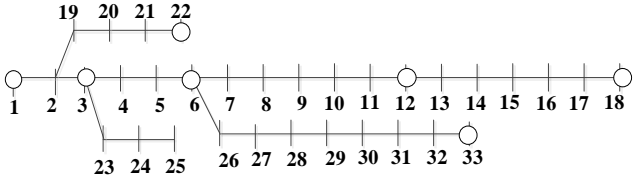

**Fig. 2** *Modified IEEE 33-bus distribution system.*

**Table 1** Coefficients of generation cost function

| DG | $c_{4,i}^{\varphi}$ | $c_{5,i}^{\varphi}$ | $c_{6,i}^{\varphi}$ |
|---|---|---|---|
| DG1-DG3 | 0.04 | 0.2 | 0 |
| DG4-DG6 | 31 | 31 | 0 |

First, we used Scenario A1, B1, C1 and D1 to analyze the three-phase DLMP and verified the effectiveness in congestions, voltage and imbalance management in ideal conditions without forecast error and uncertainty. Second, we tested the effectiveness of reactive output of DGs using Scenario D2. Third, we used Monte Carlo method to simulate the practice with forecast error and the uncertainty of prosumers' behavior using Scenario B3, C3 and D3. In each scenario, we used the LMP at the root node as flat tariff to compare the effectiveness of the proposed method with the current electricity market mechanism that charged different prosumers the same flat tariff. In response to flat tariff, each prosumer solved its own optimal problem and determined its demand and DGs' output. Then the power flows and nodal voltages in the system were obtained.

All simulations were implemented using MATLAB on a personal laptop with an Intel Core i7-7500M 2.70-GHz processor and 16 GB of RAM.

### 4.2. Three-phase DLMP and its effect on congestions, voltage and imbalance management

We considered four ideal scenarios A1, B1, C1 and D1, without forecast error and uncertainty. Here we assumed the reactive output of DGs is 0 MVar and cannot be changed. The optimal dispatch problem P2 was solved by the DSO and the three-phase DLMP was also obtained. Prosumers made response to the prices and solved their optimal problems P3 to determine their demands and DGs' outputs. Then the power flows and nodal voltages in the system were obtained.

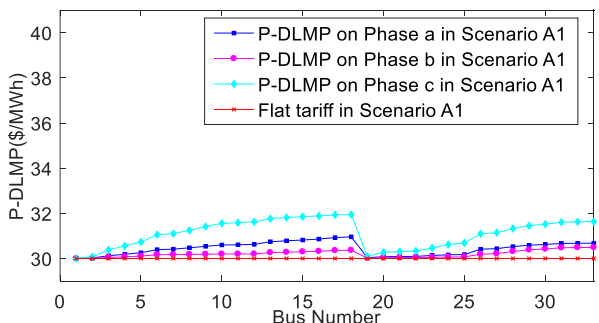

**Fig. 3** Three-phase P-DLMP and flat tariff in Scenario A1.

Scenario A1 was a base scenario that no branch power overflow, voltage violation and imbalance violation occur, with the voltage magnitude limit [0.95, 1.05] p.u., imbalance index limit 0.03, branch active power limit [-3, 3] MW. The P-DLMP and flat tariff for Scenario A1 were shown in Fig.3. It can be observed that the three-phase P-DLMP was different at different nodes and different phases, due to the different characteristic of phases. When there were no violations, the differences of price were caused by the power loss.

In Scenario B1, we tested the effectiveness of the proposed method in congestion management by setting the active power limit on branch 3 (from node 3 to node 4) as 0.5MW based on Scenario A1. Fig. 5 showed the three-phase P-DLMP and flat tariff in Scenario B1. When the congestion on branch 3- at phase c occurred, the P-DLMP of its downstream nodes increased drastically, which resulted in the decrease in the demand of corresponding nodes and the change in DGs' active output, as shown in Fig.5 and Table 2. In this way, the branch power overflow was alleviated. Fig. 6 showed the branch active power at phase c under DLMP and flat tariff in Scenario B1. The dashed line represented the transmission capacity. It can be observed that, in Scenario B1, the branch power of branch 3 at phase c was limited within its upper bound under DLMP, in this way we achieved the congestion management. We can also observe from Fig. 6 that in Scenario B1, if the prosumer made response to the flat tariff, the branch power exceeded limit, but when we used our proposed method, the violation of branch power limit was alleviated.

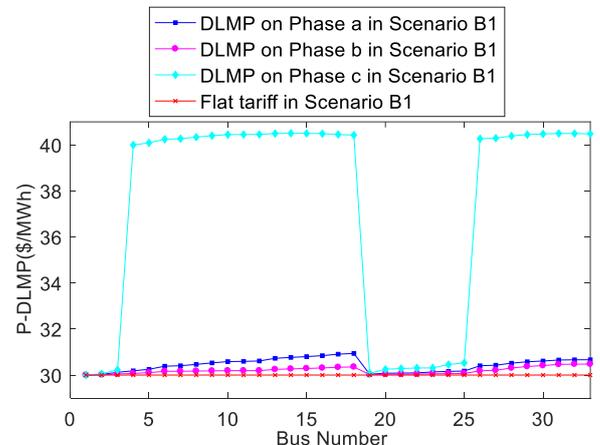

**Fig. 4** *Three-phase P-DLMP and flat tariff in Scenario B1.*

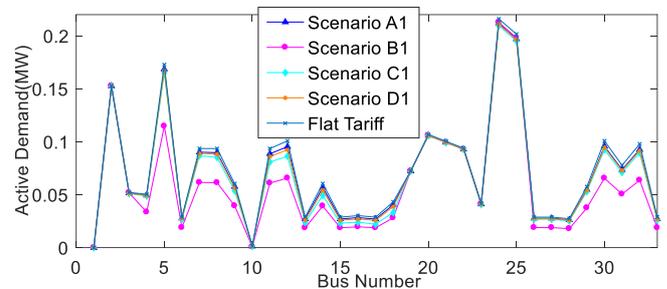

**Fig.5** *Prosumers' active demands at phase c in Scenario A1, B1, C1 and D1.*

**Table 2** Active outputs of DG4 and DG6 in scenario A1 and B1 under DLMP

| Scenario | Output of DG under DLMP (MW) | |
|---|---|---|
| | DG4 phase c | DG6 phase c |
| Scenario A1 | 0.0152 | 0.0102 |
| Scenario B1 | 0.1 | 0.1 |

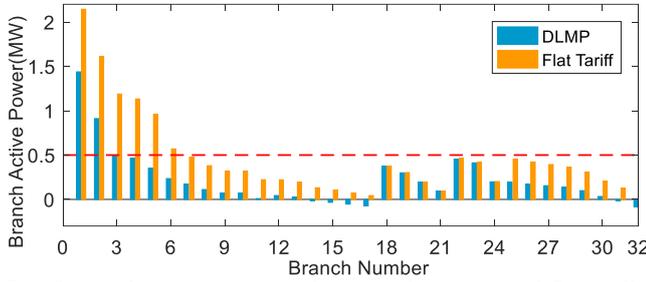

**Fig. 6** *Branch active power at phase c under DLMP and flat tariff in Scenario B1.*

In Scenario C1, we changed the voltage magnitude limit in Scenario A1 to [0.97, 1.03] p.u. to test the effectiveness of the proposed method in voltage management. Fig. 7 showed the three-phase P-DLMP and flat tariff in Scenario C1. It can be observed that, after changing the voltage limit, to maintain the voltage at phase c within the new limitation, the P-DLMP at phase c increased, inducing prosumers to decrease their demands, which was shown in Fig.5. Fig. 8 illustrated the node voltage magnitude in Scenario C1. The dashed line represented the voltage magnitude limitation. It can be observed that, the voltage exceeded limit under flat tariff, but was well maintained within the voltage limits under DLMP, then the voltage was controlled.

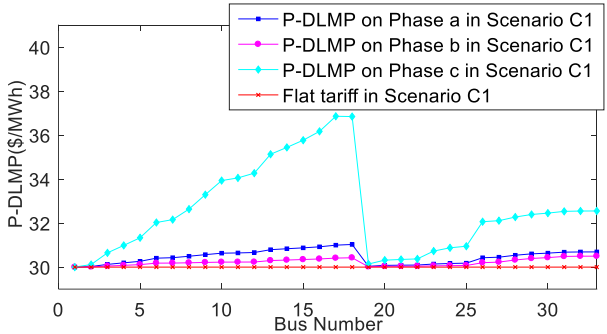

**Fig. 7** *Three-phase P-DLMP and flat tariff in Scenario C1.*

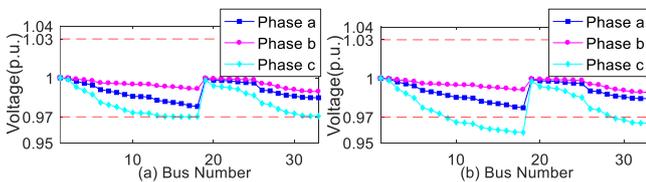

**Fig. 8** *Node voltage magnitude in Scenario C1 under DLMP and flat tariff. (a) Under DLMP in Scenario C1, (b) Under flat tariff in Scenario C1.*

In Scenario D1, the imbalance index limit $\bar{\delta}$ in Scenario A1 was changed to 0.015 to validate the effectiveness of DLMP on three-phase imbalance. Fig. 9 showed the three-phase P-DLMP and flat tariff in Scenario D1 and D2. It can be observed that, to control imbalance violation, DLMPs on phases with more load increased, inducing the change of the distribution of prosumers' demands, which was shown in Fig.5, and reduced the degree of imbalance. Fig. 10 illustrated the voltage imbalance index in Scenario

A1 and D1. The dashed line represented the imbalance limit. It can be observed that the imbalance index was kept within its limits under DLMP but exceeded its limits under flat tariff.

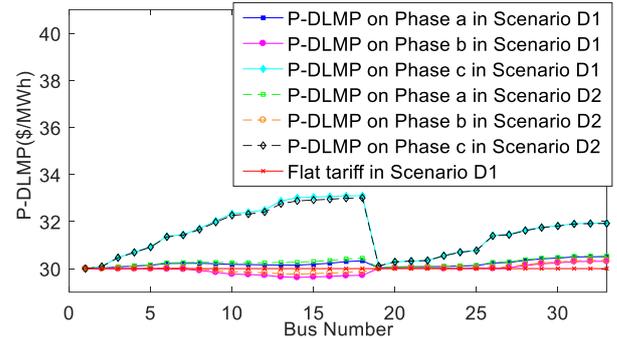

**Fig. 9** *Three-phase P-DLMP and flat tariff in Scenario D1 and D2.*

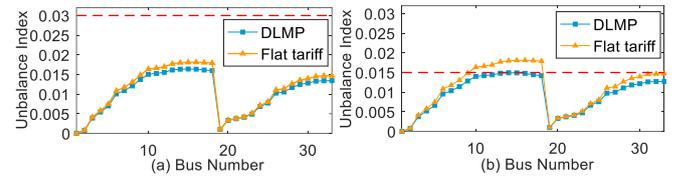

**Fig. 10** *Imbalance index in Scenario A1 and D1 under DLMP and flat tariff. (a) Scenario A1, (b) Scenario D1.*

### 4.3. Effectiveness of three-phase Q-DLMP and reactive outputs of DGs

In Scenario D2, we changed the reactive output limit of DG2, DG4, DG6 to [0, 0.1] Mvar compared with Scenario D1 to test the effectiveness of three-phase Q-DLMP and reactive output of DGs. Table 3 illustrated the three-phase Q-DLMP and the reactive outputs for DG2, DG4, DG6 in Scenario D2. It can be observed that Q-DLMPs induced the reactive output of DGs. By comparing the P-DLMPs in Scenario D1 and D2 illustrated in Fig.9, it can be observed that when DGs dispatched reactive power, the P-DLMPs changed. Table 4 was the social welfare of Scenario D1 and D2, and it showed that the social welfare increased in Scenario D2. It was because DGs' reactive outputs contributed to the voltage control process, then the active power reached a better optimal solution with less power loss, and a higher social welfare was achieved accordingly.

**Table 3** Three-phase Q-DLMP and reactive outputs of DG2, DG4, DG6 in scenario D2

| | Phase a | Phase b | Phase c |
|---|---|---|---|
| Q-DLMP for DG2. ($/MVarh) | 2.1901 | 2.4534 | 5.8152 |
| Reactive output of DG2.(MVar) | 0.1 | 0.1 | 0.1 |
| Q-DLMP for DG4. ($/MVarh) | 3.9985 | 4.1101 | 13.4571 |
| Reactive output of DG4.(MVar) | 0.1 | 0.1 | 0.1 |
| Q-DLMP for DG6. ($/MVarh) | 3.6327 | 5.0894 | 6.8556 |
| Reactive output of DG6.(MVar) | 0.1 | 0.1 | 0.1 |

**Table 4** Social welfare of scenario D1 and D2

| | Scenario D1 | Scenario D2 |
|---|---|---|
| Social Welfare ($) | 82.626 | 83.215 |

### 4.4. Analysis of the proposed method considering forecast error and uncertainty

In practice, the forecast error and the uncertainty of prosumers' behavior are always involved. To simulate the practice, we formulated the Scenario B3, C3 and D3 based on Scenario B1, C1 and D1. We assumed that the forecast error was normally distributed. The mean value was 0, and the standard deviation was 1% of the predicted value. The uncertainty of prosumers' utilities was simulated as the utility function coefficient obeyed normal distribution. The mean value was original coefficient, and the standard deviation was 0，1%，2%，3%，4%，5% of the original coefficient, respectively. The Monte Carlo method was used with 500 tests to achieve the result.

Table 5 Average Over-limit Branch Active Power, Voltage Magnitude, Imbalance Index in Scenario B3, C3 and D3 under DLMP and Flat Tariff

| Standard deviation of the prosumers' utility function coefficients | Average over-limit Branch Active Power in Scenario B3(MW) | | Average over-limit Voltage Magnitude in Scenario C3 (p.u.) | | Average over-limit Imbalance Index in Scenario D3 | |
|---|---|---|---|---|---|---|
| | DLMP | Flat Tariff | DLMP | Flat Tariff | DLMP | Flat Tariff |
| 0% | 0.0018 | 0.688 | 1.53e-4 | 0.0940 | 2.65e-5 | 0.0228 |
| 1% | 0.0033 | 0.688 | 3.39e-4 | 0.0939 | 1.28e-4 | 0.0229 |
| 2% | 0.0063 | 0.688 | 6.45e-4 | 0.0938 | 3.72e-4 | 0.0230 |
| 3% | 0.0093 | 0.687 | 1.04e-3 | 0.0937 | 6.61e-4 | 0.0231 |
| 4% | 0.0116 | 0.687 | 1.54e-3 | 0.0936 | 9.43e-4 | 0.0230 |
| 5% | 0.0171 | 0.689 | 2.23e-3 | 0.0937 | 1.34e-3 | 0.0229 |

Table 5 showed the average over-limit branch active power, average over-limit voltage magnitude and average over-limit imbalance index in Scenario B3, C3 and D3 under DLMP and flat tariff, respectively. It can be observed that when there was only forecast error in the system, which meant the utility uncertainty standard deviation was 0, the violation in branch power, voltage and imbalance under the proposed DLMP method was much lower than that under the flat tariff. With the increase of prosumers' uncertainty in utility function, the violation under DLMP increased, but the corresponding parameters were still much lower than those under the flat tariff. Which indicated the proposed method achieved a better management in congestions, voltage and imbalance, although there were forecast errors and uncertainty in the system. It can be concluded that by using the proposed method, serious violations of the operation limits will not occur.

## 5. Conclusion

This paper proposes an optimal dispatch and pricing method to alleviate imbalance, congestions and voltage violations based on three-phase DLMP. A three-phase convex QP OPF model considering the limits on the degree of three-phase imbalances, voltage magnitudes and branch power flows is presented for DSO. Three-phase Q-DLMP is determined for reactive power. DSO solves the OPF model and calculates three-phase DLMPs for active power and reactive power. Because the DSO's problem and the prosumers' problems are equivalent, prosumers will voluntarily follow the instructions of DSO and then branch power overflows, voltage violations and three-phase imbalance will be alleviated. The proposed method was tested on a modified three-phase imbalanced IEEE 33-bus distribution system. Simulation results indicate the three-phase DLMP will be different at different locations and phases. Under the incentive of the different DLMP prices, prosumers will adjust their demands and DGs' outputs, then the imbalance, congestions and voltage violations will be alleviated. In addition, DGs' reactive output will contribute to the management to make the social welfare increase. Meanwhile, the test results demonstrate that the proposed method will achieve a better management compared with current mechanism using flat tariff signal, although there are forecast errors and uncertainty in the system.

## 6. Acknowledgments

This work was supported in part by the National Science Foundation of China (51725703).

## 8. Appendix

In this section, we elaborate the derivation processes of (15).

According to

$$\left(V_i^a\right)^2 + \left(V_i^b\right)^2 \geq 2V_i^a V_i^b,$$
$$\left(V_i^a\right)^2 + \left(V_i^c\right)^2 \geq 2V_i^a V_i^c, \quad (36)$$
$$\left(V_i^b\right)^2 + \left(V_i^c\right)^2 \geq 2V_i^b V_i^c,$$
$$V_i^a, V_i^b, V_i^c \approx 1,$$

we have

$$\left(\frac{V_i^a}{V_i^a + V_i^b + V_i^c}\right)^2$$
$$= \frac{\left(V_i^a\right)^2}{\left(V_i^a\right)^2 + \left(V_i^b\right)^2 + \left(V_i^c\right)^2 + 2V_i^a V_i^b + 2V_i^b V_i^c + 2V_i^a V_i^c} \quad (37)$$
$$\approx \frac{\left(V_i^a\right)^2}{3\left(V_i^a\right)^2 + 3\left(V_i^b\right)^2 + 3\left(V_i^c\right)^2}.$$

According to (1) and (14), we have

$$\frac{1-\bar{\delta}}{3} \leq \frac{V_i^\varphi}{\sum V_i^\varphi} \leq \frac{\bar{\delta}+1}{3}, \varphi = a,b,c \quad . \quad (38)$$

According to (37) and (38), we get (15).